\documentclass[10pt,conference]{IEEEtran}
\IEEEoverridecommandlockouts

\usepackage{cite}
\usepackage{amsmath,amssymb,amsfonts}
\usepackage{algorithmic}
\usepackage{graphicx}
\usepackage{textcomp}
\usepackage{xcolor}
\usepackage{hyperref}
\usepackage[capitalise]{cleveref}
\usepackage[acronym]{glossaries}
\usepackage{xspace}
\usepackage{url}
\usepackage{subcaption}
\usepackage[colorinlistoftodos,prependcaption,textsize=tiny]{todonotes}
\usepackage{balance}
\usepackage{float}

\def\BibTeX{{\rm B\kern-.05em{\sc i\kern-.025em b}\kern-.08em
		T\kern-.1667em\lower.7ex\hbox{E}\kern-.125emX}}

\newcommand{\toolname}{\emph{Code Critters}\xspace}

\newcommand{\school}{\emph{Maristengymnasium Fürstenzell}\xspace}

\newcommand{\summary}[2]{%
	\vspace{-0.2cm}%
	\begin{center}%
		\colorbox{gray!20}{%
			\parbox{\linewidth}{%
				\textbf{\textsf{Summary (\textit{#1})}:}~%
				#2%
			}%
		}%
	\end{center}%
}

\begin{document}
	
	\title{Teaching Loop Testing to Young Learners with the Code Critters Mutation Testing Game}
	\author{\IEEEauthorblockN{Philipp Straubinger}
			\IEEEauthorblockA{\textit{University of Passau} \\
					Passau, Germany}
			\and
			\IEEEauthorblockN{Lena Bloch}
			\IEEEauthorblockA{\textit{University of Passau} \\
				Passau, Germany}
				\and
				\IEEEauthorblockN{Gordon Fraser}
				\IEEEauthorblockA{\textit{University of Passau} \\
						Passau, Germany}
			}
		
		\maketitle
		
		\begin{abstract}
			Serious games can teach essential coding and testing concepts even
			to younger audiences.  In the \toolname game critters execute short
			snippets of block-based code while traversing the game map, and
			players position magical portals (akin to test oracles) at locations
			(akin to test inputs) to distinguish between critters
			executing correct code from those who execute faulty code. However,
			this adaptation of the tower defense genre limits code under test to
			basic sequences and branches, and excludes the fundamental
			programming concept of loops.
			To address this limitation, in this paper we introduce an entirely
			new game concept integrated into the \toolname storyline, tasking
			players to test the behavior of critters collecting ingredients for a healing potion using loop-based
			recipes at a second-stage level.
			In a study involving 29 secondary school students, we observed
			active engagement with these new loop-integrated levels. The results
			highlight challenges the students face, which can inform
			future strategies for improving coding and testing education.
		\end{abstract}
		
		
		\section{Introduction}
		Coding is not only the foundation of software engineering, but it is also becoming a prominent part of school curricula worldwide, even for young learners. However, learning to code presents challenges, often hindered by students' misconceptions about fundamental coding concepts like loops and variables~\cite{qian2017students,sorva2012visual}. 
To support students in overcoming these difficulties, testing has been suggested as an effective approach to enhance their understanding of programming principles~\cite{denny2019closer,prather2018metacognitive,wrenn2019executable,prasad2023conceptual}. Research indicates that introducing testing concepts early can significantly improve students' grasp of coding practices and their ability to create more reliable programs~\cite{DBLP:conf/acse/Carrington97,jones2001experiential,DBLP:conf/iticse/MarreroS05}. 
Unfortunately, engaging younger learners in abstract concepts like software testing remains a major challenge~\cite{DBLP:conf/iticse/Luxton-ReillySA18a,DBLP:conf/acse/Carrington97,jones2001experiential,DBLP:conf/iticse/MarreroS05}.

Gamification has become a promising strategy to tackle engagement issues in education~\cite{DBLP:conf/mindtrek/DeterdingDKN11,DBLP:journals/computers/NardoFFMMS24}. Specifically, serious games designed for learning have proven effective in teaching various programming and debugging skills~\cite{DBLP:conf/icer/MiljanovicB17,DBLP:conf/its/MuratetDTV12}. 
For software testing, games like Code Defenders~\cite{DBLP:conf/sigcse/FraserGKR19} and Code Immunity Boost~\cite{hsueh2023design} have shown the benefits of gamifying testing concepts to engage learners. However, these games are primarily aimed at students with more advanced mechanical, reading, and programming skills, which limits their accessibility for younger or less experienced learners~\cite{DBLP:journals/jss/GarousiRLA20}.

The serious game \toolname~\cite{DBLP:conf/icst/StraubingerCF23,DBLP:conf/icst/StraubingerBF24} introduces a novel approach designed to bridge the gap between the challenges of engaging with testing concepts through gamification and the abilities of younger learners. Like advanced testing games such as Code Defenders and Code Immunity Boost, \toolname is based on the concept of mutation testing, where artificially seeded bugs are used to assess test effectiveness and inspire new tests. In \toolname, this concept is subtly woven into a creative storyline where critters run short code snippets as they move through a landscape of tiles, and players must create portals to separate critters running correct code from those running mutated code.
Unlike other testing games, \toolname is specifically designed to accommodate the needs of younger learners by utilizing block-based programming, which removes the syntax-related challenges of traditional text-based coding~\cite{maloney2010scratch}. This has been shown to be effective in engaging students in testing activities~\cite{DBLP:conf/icst/StraubingerBF24}.

Although \toolname has proven effective, the approach of having critters execute code snippets during each step of their journey inherently limits the range of programming concepts that can be addressed. While \toolname successfully covers basic sequences, conditional branches, and variables, it notably lacks the integration of loops into its gameplay. Loops are a crucial programming concept that children often find challenging to grasp~\cite{DBLP:journals/ijcci/GomesFT18,DBLP:conf/sigcse/GroverB17,bentz2023novice} and for which they are especially prone for errors~\cite{DBLP:journals/tse/Ntafos88}. 

To address this gap, in this paper we present a new game concept that enhances the \toolname serious game by making loops and the testing of loop-based programs a central element of its mutation testing-based gameplay. 
This updated concept builds on the existing \toolname storyline by introducing second-stage levels: after rescuing the healthy critters through the original gameplay, players use loop-based recipes to create healing potions. In this stage, players must differentiate between critters following the correct recipe and those following mutated versions, requiring them to understand loops and assert the expected behavior effectively.

Overall, the contributions of this paper are as follows:
\begin{itemize}
\item We introduce a novel game concept that combines mutation testing
  with loop testing.
\item We integrate the loop testing game concept into the \toolname
  narrative and gameplay.
\item We implement the new game concept as an extension of \toolname,
  and provide multiple playable loop-levels.
\item We empirically evaluate \toolname and the new loop levels
  involving 29 school students.
\item We discuss the challenges the children encountered and suggest
  potential solutions to address them.
\end{itemize}

Our study involving secondary school students revealed that these new
levels promote active engagement, though they also present unique
challenges in grasping loop concepts. These findings suggest potential
strategies for enhancing the way programming and testing are taught to
younger learners.


		\section{Background}

Key programming constructs—such as sequences, conditionals, and
especially loops or repetitions are fundamental to learning
programming. However, learning to program remains a complex task for
young learners, who may have misconceptions about fundamental concepts
like loops~\cite{qian2017students,sorva2012visual}, which are known to
be one of the most challenging concepts for children to
grasp~\cite{DBLP:journals/ijcci/GomesFT18,DBLP:conf/sigcse/GroverB17}. Research
has shown that concrete, hands-on experiences can significantly aid
learners in overcoming these difficulties and developing abstract
thinking
skills~\cite{DBLP:journals/ijcci/GomesFT18,DBLP:conf/sigcse/GroverB17,DBLP:journals/eait/MladenovicBZ18}. To
ensure learners are engaged in these experiences, gamification has
been shown to be an effective solution.

\subsection{Educational  Testing Games}

Gamification integrates game-like elements—such as leaderboards,
points, and challenges—into non-game
settings~\cite{DBLP:conf/mindtrek/DeterdingDKN11}.
Serious games take this approach further by explicitly aiming to
educate, train, or simulate real-world tasks with real gameplay. By engaging players
with embedded learning experiences, such games allow them to acquire
knowledge or skills without feeling as though they are in an
educational setting~\cite{DBLP:conf/chi/RaybournB05}. Consequently,
many educational games focus on various aspects of
programming~\cite{DBLP:journals/bjet/LindbergLH19,DBLP:journals/ijcci/MacridesMA22,DBLP:journals/ijcci/GomesFT18}. The
loop programming concept has also been the focus of gamification
strategies and serious games to aid
learning~\cite{shorn2018teaching,DBLP:conf/dac/BoroujerdianGCP21,zhao2019improving,makri2019computer}.

Although testing has been proposed as a means of supporting student
understanding~\cite{denny2019closer,prather2018metacognitive,wrenn2019executable,prasad2023conceptual},
and introducing testing concepts early has been shown to improve
students' understanding of coding practices and their ability to write
robust programs~\cite{DBLP:conf/acse/Carrington97,
	jones2001experiential,DBLP:conf/iticse/MarreroS05}, it remains
underemphasized in programming
instruction~\cite{DBLP:journals/jss/GarousiRLA20}. Serious gaming has
also seen only limited applications related to software
testing~\cite{DBLP:conf/icer/MiljanovicB17,
	DBLP:conf/icse/PrasetyaLMTBEKM19, DBLP:conf/fie/ToledoLS22}.

One successful approach to gamifying testing concepts is to use
mutation testing, as exemplified by games like Code
Defenders~\cite{DBLP:conf/sigcse/FraserGKR19}, Code Immunity
Boost~\cite{hsueh2023design} or the Testing
Game~\cite{DBLP:conf/fie/ValleTBM17}. In mutation testing, artificial
defects (or mutants) are deliberately introduced into the code to
assess the robustness of existing tests~\cite{5487526}. The process
involves creating slight variations of the code under test and running
the available test suite against them. When a test fails, it indicates
the mutant has been detected or \emph{killed}; if the mutant survives,
it highlights potential gaps in the test coverage and assertion
quality. Games like Code Defenders~\cite{DBLP:conf/sigcse/FraserGKR19}
gamify this process by allowing ``attackers'' to generate mutants,
while ``defenders'' work to identify and eliminate these defects by
writing tests.

Existing testing games typically require advanced programming
knowledge, limiting their accessibility to higher education
audiences. To engage younger or less experienced learners, testing
needs to be introduced earlier in programming curricula and presented
in a more approachable manner. A promising solution to this challenge
lies in the use of block-based programming environments, such as
Scratch~\cite{maloney2010scratch}, which has been successful in
making programming accessible to beginners. In these environments,
learners construct programs by dragging and connecting visual blocks
of code, bypassing the complexity of traditional textual
programming. This intuitive method has been shown to lower the
barriers to entry for novice programmers, enabling them to create
games and other programs
quickly~\cite{DBLP:journals/cacm/BauGKST17}. This concept is used by
the Code Critters mutation testing
game~\cite{10675860,DBLP:conf/icst/StraubingerCF23}, a block-based
approach to introduce software testing concepts to younger learners,
making testing more accessible.

\subsection{The Code Critters Game}

\label{sec:codecritters}

\toolname is an educational browser game that aims to make software
testing fun and engaging for young learners. It blends the classic
Tower Defense game style with lessons on block-based programming and
software testing, all spread across ten levels.
%
The game focuses on critters, who are a humanoid species that have
long lived peacefully in the deep, mysterious forests of an isolated
land. Their harmonious existence is disrupted when a sudden outbreak
of an unknown disease spreads through their colony. As more critters
become infected, they begin to show unusual and disturbing behavior,
threatening the colony's peace. In response, the remaining critters
are forced to abandon their city and seek refuge in a safe tower deep
within the forest.

\begin{figure}
	\centering
	\includegraphics[width=\linewidth]{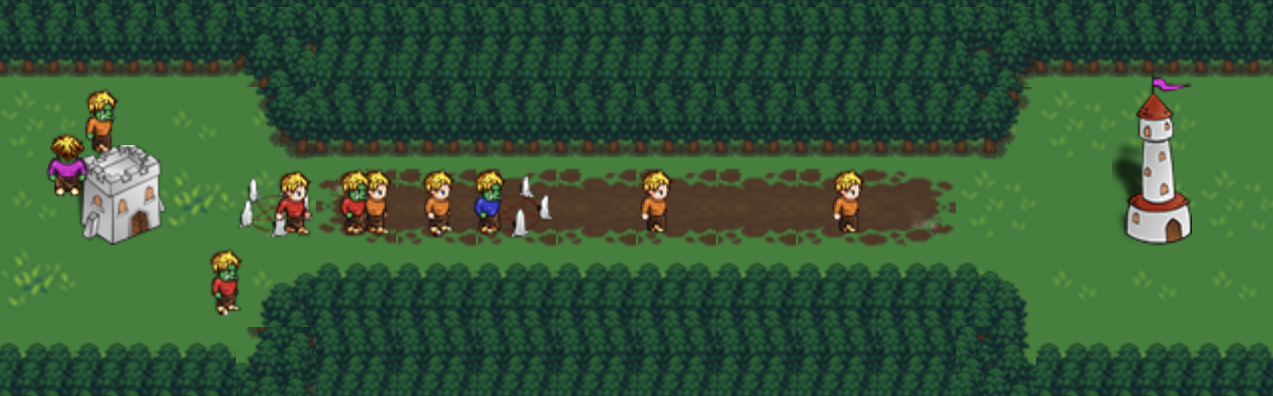}
	\caption{Gameboard of \toolname during gameplay}
	\label{fig:baseGameplay}
\end{figure}

The initial levels of \toolname portray this evacuation, as illustrated in \cref{fig:baseGameplay}. Both infected and healthy critters attempt to flee their village and reach the safe tower. The player's objective is to ensure that only the healthy critters make it to safety, using magical portals to keep the infected ones away.

\subsubsection{Game Concept}

\begin{figure}
	\centering
	\includegraphics[width=\linewidth]{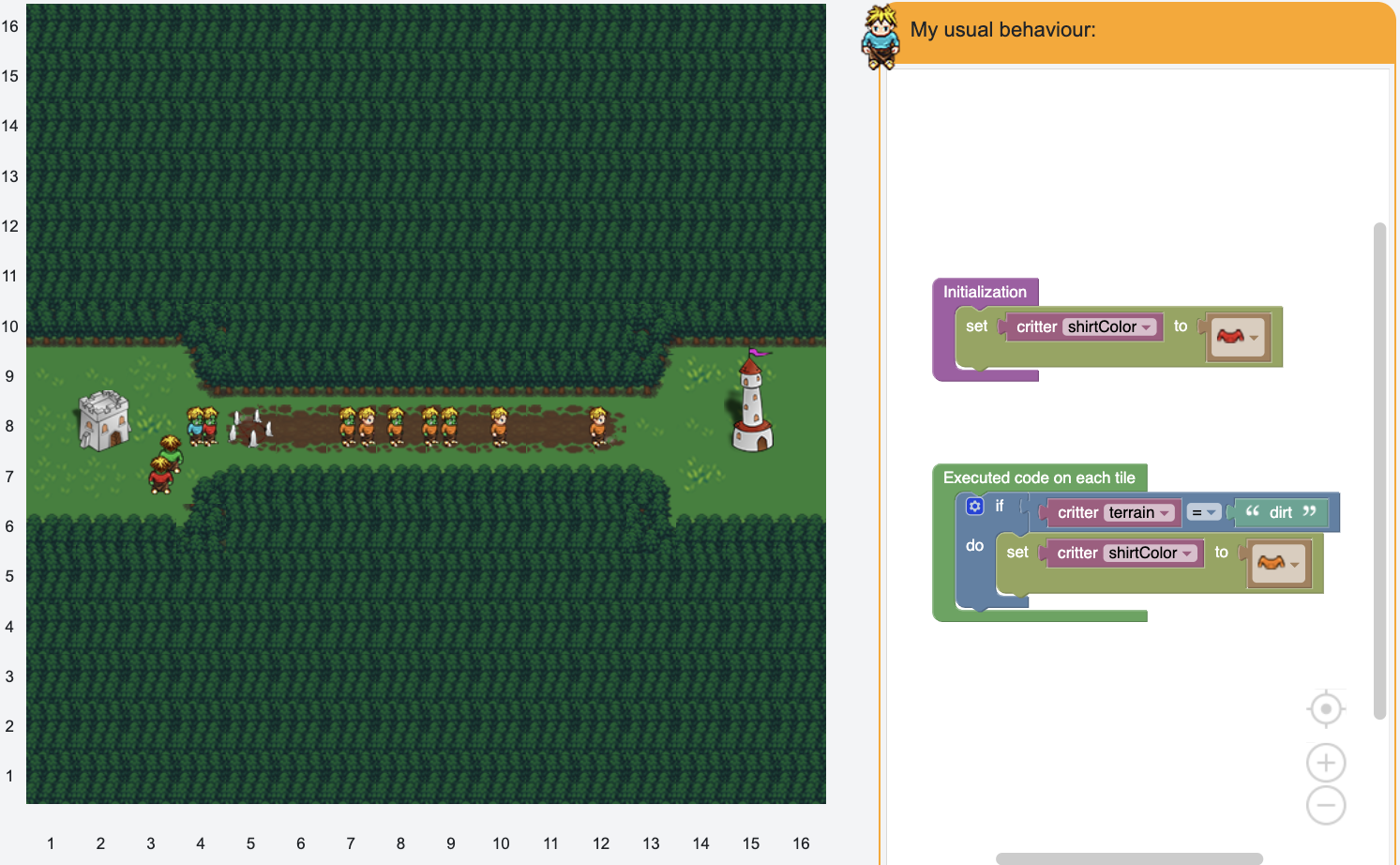}
	\caption{Game screen of base level 1}
	\label{fig:gameScreen}
\end{figure}

As the critters make their way from the village to the tower, the
player's task is to identify and separate the healthy critters from
the infected ones (mutants) to protect the tower. The infected
critters can be easily recognized by their mutated, green appearance
(\cref{fig:baseGameplay}). However, the most important difference
between healthy critters and mutants lies in their behavior. This
behavior is referred to as the Critter Under Test (CUT) and consists
of a series of code instructions that guide the critters' actions as
they move.
The orange-framed box on the right side of \cref{fig:gameScreen} shows
the Critter Under Test (CUT) for Level 1, alongside the corresponding
game board on the left. 
\toolname utilizes the
Blockly\footnote{https://developers.google.com/blockly} library, which
provides visual, block-based programming instead of traditional
text-based code, making it more accessible for beginners. With each
level, the CUT's complexity increases, offering progressively more
challenging gameplay.

The functions within the CUT are similar to
methods in an object-oriented class, defining the critters' attributes
and behavior.
The \textit{Initialization} section (\cref{fig:gameScreen}) sets the initial attributes of the critters as they leave their village. The \textit{Executed code on each tile} section outlines the critter's behavior as it moves through the game, with specific instructions that are carried out on each tile it steps on. Unlike healthy critters, mutants deviate from the defined behavior in at least one way.

In the Level 1 example shown in \cref{fig:gameScreen}, a healthy critter starts wearing a red shirt and changes to an orange shirt when moving onto a dirt tile. In contrast, the altered code of a mutant might cause it to start with a blue shirt instead of red or change to a pink shirt instead of orange when stepping on dirt.

The game board that the critters navigate is made up of a 16 by 16 grid of tiles with different terrains: grass, dirt, ice, water, or wood. Critters can only walk on grass, dirt, and ice. The game board is displayed on the left-hand side of \cref{fig:gameScreen}, with one portal placed along the path.

\subsubsection{Gameplay}

\begin{figure}
	\centering
	\includegraphics[width=\linewidth]{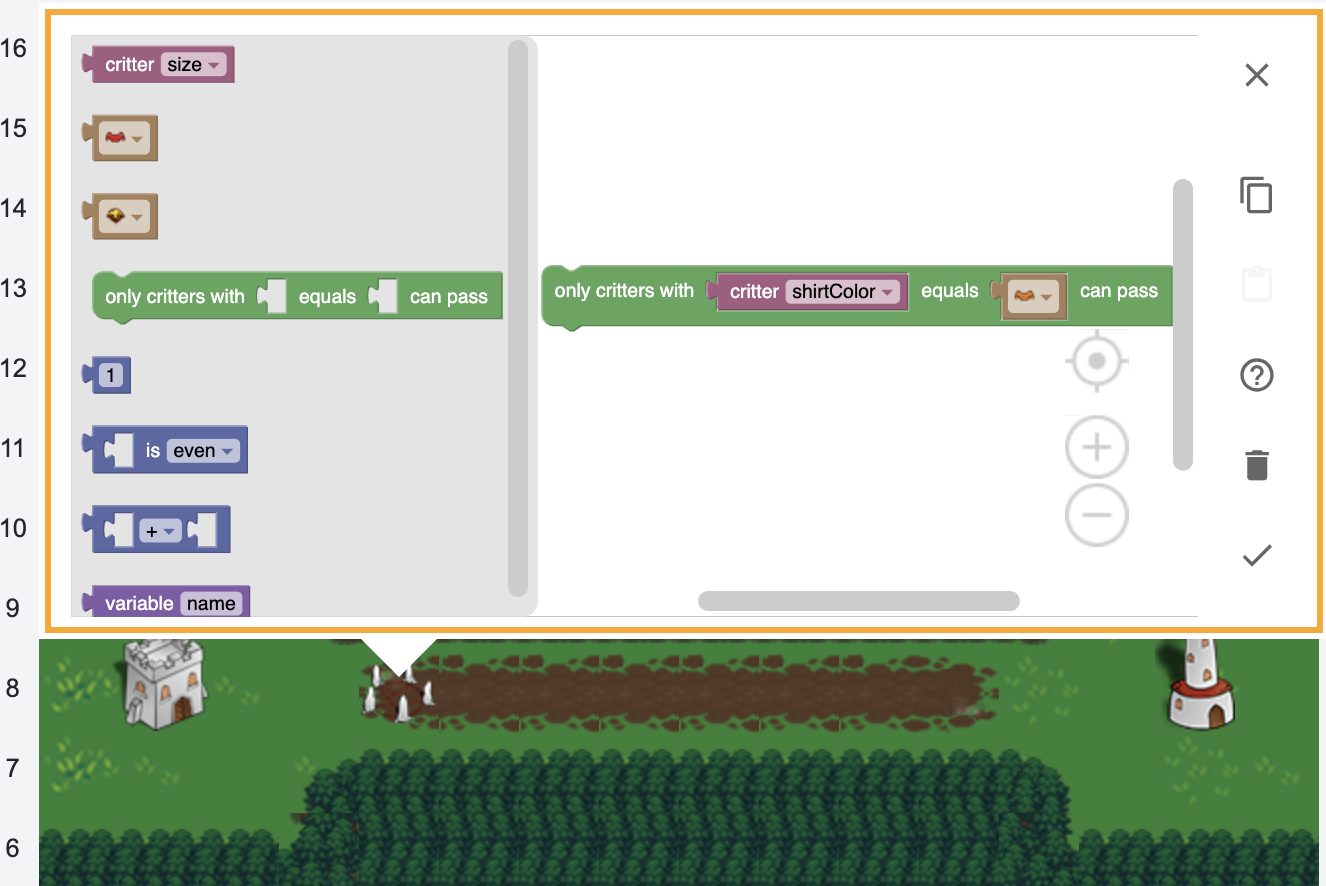}
	\caption{Opened portal of base level 1}
	\label{fig:gameWithPortal}
\end{figure}

The game board features the village where the critters start walking, the tower as their destination, and at least one path connecting them through the forest. The player's goal is to strategically place portals along the critters' path, allowing only the uninfected ones to pass. Mutants, on the other hand, are teleported by these portals to a safe location until a cure is discovered.

Portals act as block-based tests that help distinguish between mutant and healthy behavior. To ensure that no mutants reach the tower, the player must carefully check the CUT. \Cref{fig:gameWithPortal} illustrates the code for a portal placed on the first dirt tile. This test runs when a critter steps on that tile, verifying the expected behavior. Similar to traditional assertions, the \textit{only critters with ... equals ... can pass} block ensures that critters have the correct attributes when they reach specific fields.

Once the game begins, critters leave the village and move toward the tower, stepping onto tiles with portals along the way. As they walk across portals, these execute their checks, e.g., for correct shirt color transitions in \cref{fig:gameWithPortal}, allowing passage \textit{only} to critters dressed in orange. If a mutant is detected, it is teleported away, visualized by the captured mutant flying across the screen to a holding area. Note that players can only see the code of correct critters while playing, while mutations are only shown after level completion.

The game ends once the last critter either reaches the tower or is intercepted by a portal. The player then receives a score based on the percentage of identified mutants and the number of healthy critters that reach the tower. Achieving a perfect score of 1000, or three stars, requires correctly capturing all mutants while letting all healthy critters pass. Players can earn up to an additional 10\% bonus for quickly installing the portals, bringing the maximum score to 1100.

\begin{figure}
	\centering
	\includegraphics[width=0.8\linewidth]{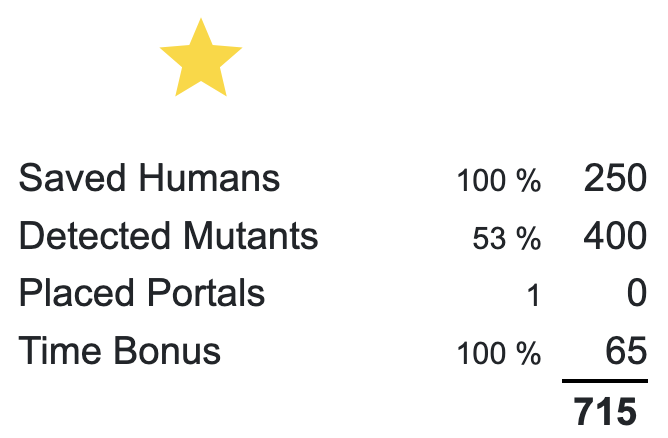}
	\caption{Scoreboard after finishing a base level}
	\label{fig:baseScoreDialog}
\end{figure}

\Cref{fig:baseScoreDialog} shows the score dialog after completing Level 1 using the portal setup from \cref{fig:gameWithPortal}. In this example, the player did not achieve a perfect score because not all mutants were successfully captured. To catch the remaining mutants and earn a full score, an additional test (i.e., portal) is needed to ensure that critters start their path with the correct red shirt.

\begin{figure*}
	\centering
	\includegraphics[width=0.8\linewidth]{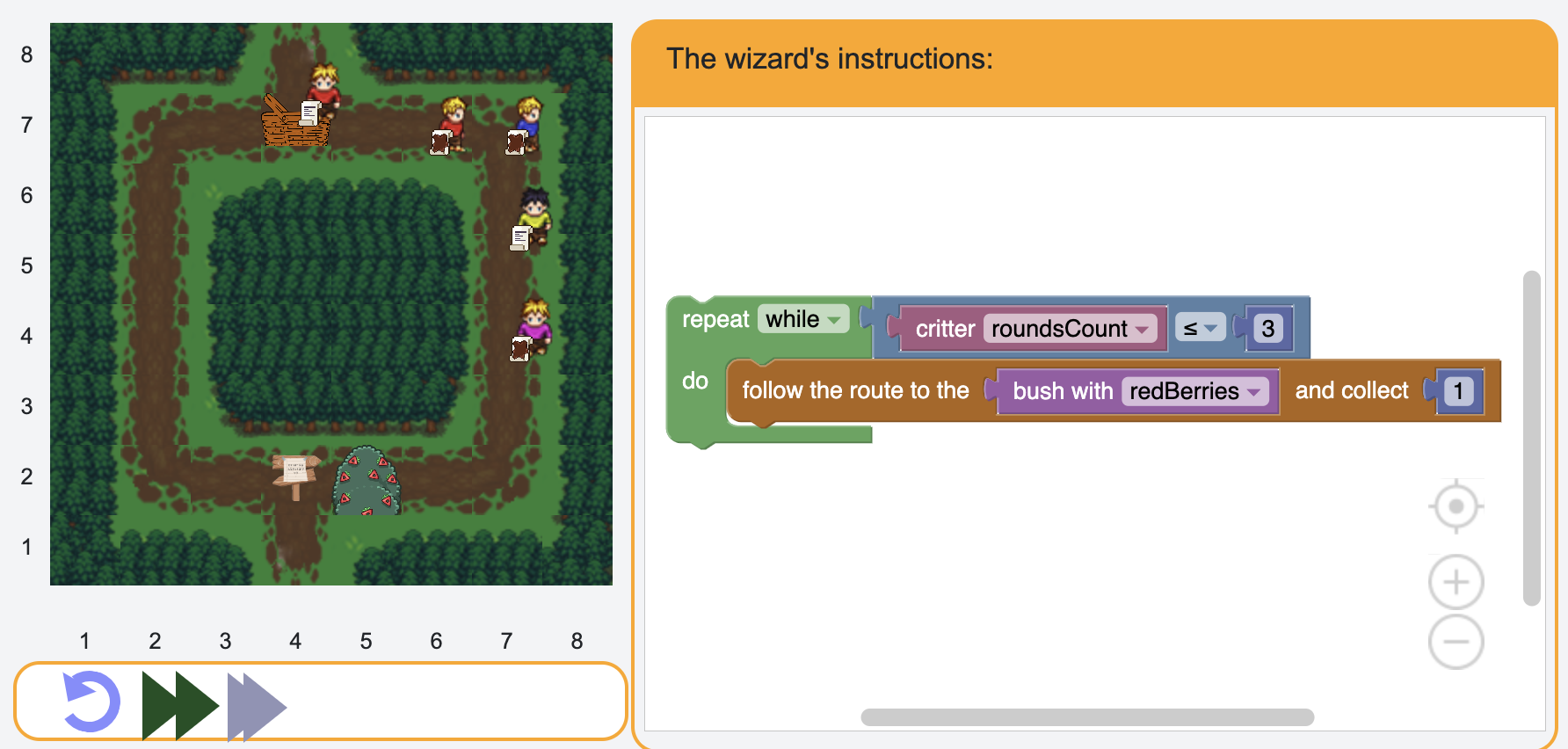}
	\caption{Game screen of loop level 1}
	\label{fig:loopLevel1screen}
\end{figure*}


		\section{Loop Testing with Code Critters}
		\begin{figure}
	\centering
	\includegraphics[width=0.8\linewidth]{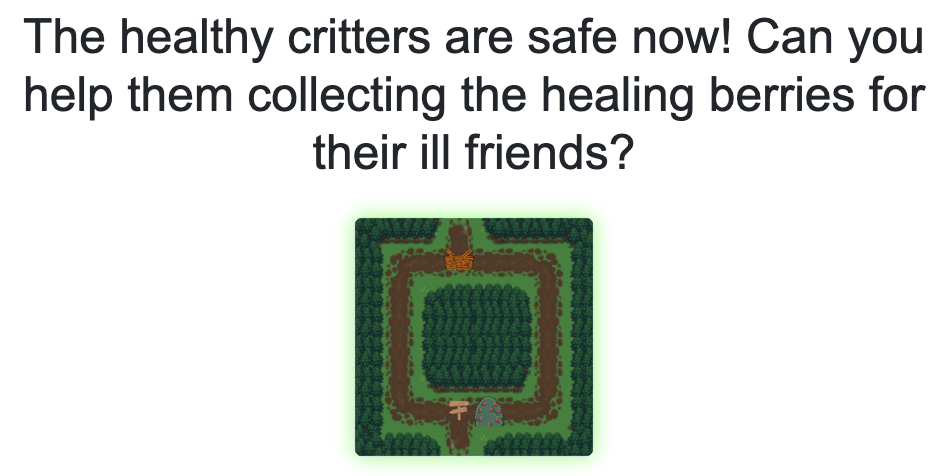}
	\caption{Second part of the scoreboard after a successful game}
	\label{fig:baseScoreDialogUnlockedSecond}
\end{figure}

In order to integrate loop testing into \toolname, we extend existing
base levels of \toolname, as described in \cref{sec:codecritters},
with new second-stage levels. Like the base levels, the second stages
also use the concept of mutation testing but place a stronger emphasis
on a previously overlooked topic: loops. These loop levels are not
available from the start; they must be unlocked by scoring at least
800 points in the corresponding base levels. This design encourages
players to first develop their skills and deepen their understanding
of the basic game concepts. \Cref{fig:baseScoreDialogUnlockedSecond}
displays the conclusion dialog shown to players who completed a base
level with a perfect score earning three stars and thus enabling the
second stage.

\subsection{Story}

The healthy critters are determined to find a way to cure their sick companions, so they seek help from a friendly wizard. The wizard agrees to assist them but needs the critters to gather magical berries to brew a healing potion. Armed with a magical recipe detailing how to collect the berries, along with a basket and supplies provided by the wizard, the critters set off on their journey. However, the paths to the magical berries have been long neglected, making it difficult for the critters to keep their recipes clean and unspoiled.

\subsection{Gameplay}

\Cref{fig:loopLevel1screen} shows the game screen during gameplay of the first loop level. Following the basic structure of the base levels, the game board appears on the left, with the corresponding Recipe Under Test (RUT) displayed on the right. The game board consists of an 8-by-8 grid of tiles, featuring a cyclic path with a collection basket at its start, at least one berry bush, and a signpost along the way. 

Since critters cannot carry all the required berries at once, they must gather them in multiple rounds, gradually filling the wizard's basket. The RUT is structured as a loop, representing the critters' repeated trips to gather berries, with instructions specifying how to collect them. For example, when following the instructions in \cref{fig:loopLevel1screen}, a critter completes three rounds and picks exactly one red berry in each round. If the recipe is smudged or corrupted (i.e., a mutant), however, a critter might for example collect two berries instead of one, deviating from the valid instructions. Mutants must be sent back to the wizard to receive a clean recipe. They can be visually identified by the dirt-smudged piece of paper they carry.

The objective of this level, similar to the base levels, is to identify these mutants among the berry collectors using assertion tests. Instead of portals, signposts serve as checkpoints to distinguish and send back the critters carrying corrupted recipes. The signposts can be filled with the same familiar block-based code as in the base levels to assert relevant conditions. This toolbox offers a wider range of options, including if-else blocks, to handle the more complex tests required for loop-based scenarios.

\begin{figure}
	\centering
	\includegraphics[width=\linewidth]{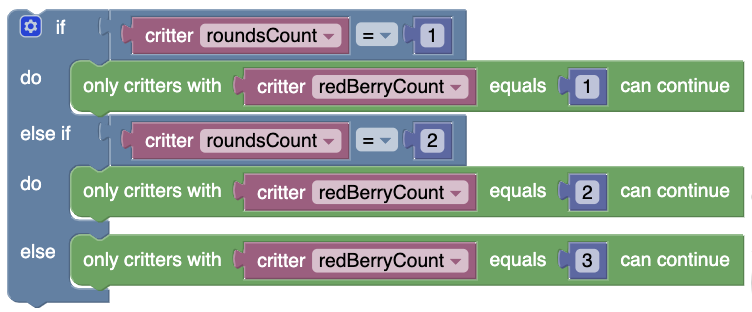}
	\caption{Correct but long test for loop level 1}
	\label{fig:validTestLong}
\end{figure}

\begin{figure}
	\centering
	\includegraphics[width=\linewidth]{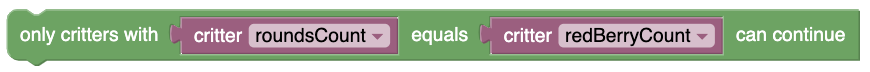}
	\caption{Correct but short and efficient test for loop level 1}
	\label{fig:validTestShort}
\end{figure}

These blocks can be used to create customized tests that differentiate between valid and corrupted berry collectors, allowing for multiple approaches. For example, two versions of a correct test for level 1 (\cref{fig:loopLevel1screen}) are illustrated in \cref{fig:validTestLong} and \cref{fig:validTestShort}. Both tests are equally effective in verifying the correct berry collection during each round—whether by using if-conditions for an explicit check in each round or by expecting the number of berries to match the number of rounds completed. Although technically signposts are similar to the test cases created in the base levels, conceptually loop tests are similar to loop invariants~\cite{DBLP:journals/csur/FuriaMV14}. 

Once the signpost is set up, the game can be started, and the critters begin their rounds of gathering berries. Each time a critter passes the signpost, the test is executed. If the test fails, indicating that the critter has a corrupted recipe, the critter exits at the crossing to return to the wizard for a new recipe. On the other hand, critters who pass the test have the correct recipe and continue with their gathering mission. The game ends when the last critter either completes its task or is sent back, followed by a dialog displaying the player's final score.

\begin{figure}[t]
	\centering
	\includegraphics[width=\linewidth]{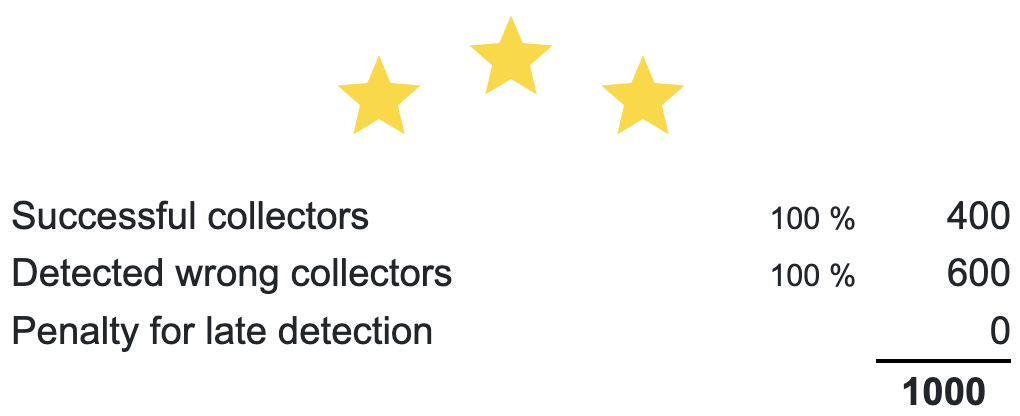}
	\caption{Scoreboard after finishing a loop level}
	\label{fig:scoreLoop}
\end{figure}

\Cref{fig:scoreLoop} shows the conclusion dialog after successfully completing a loop level. In each loop level, there are ten critters on their journey, with a varying number of mutants to detect. The scoring system is similar to that of the base levels, with the player's score primarily based on the percentage of correctly identified mutants and valid collectors. The maximum score is 1000 points or three stars.

A unique feature of the loop levels is the penalty for late detection, which emphasizes the importance of early error identification in repetitive tasks. Points are deducted if mutants remain in the game for longer than they should. In other words, if a mutant is not caught in the first round where its mutation affects the outcome, each additional round it remains undetected results in a penalty of 25 points.

\subsection{Progression}

\begin{figure}
	\centering
	\includegraphics[width=\linewidth]{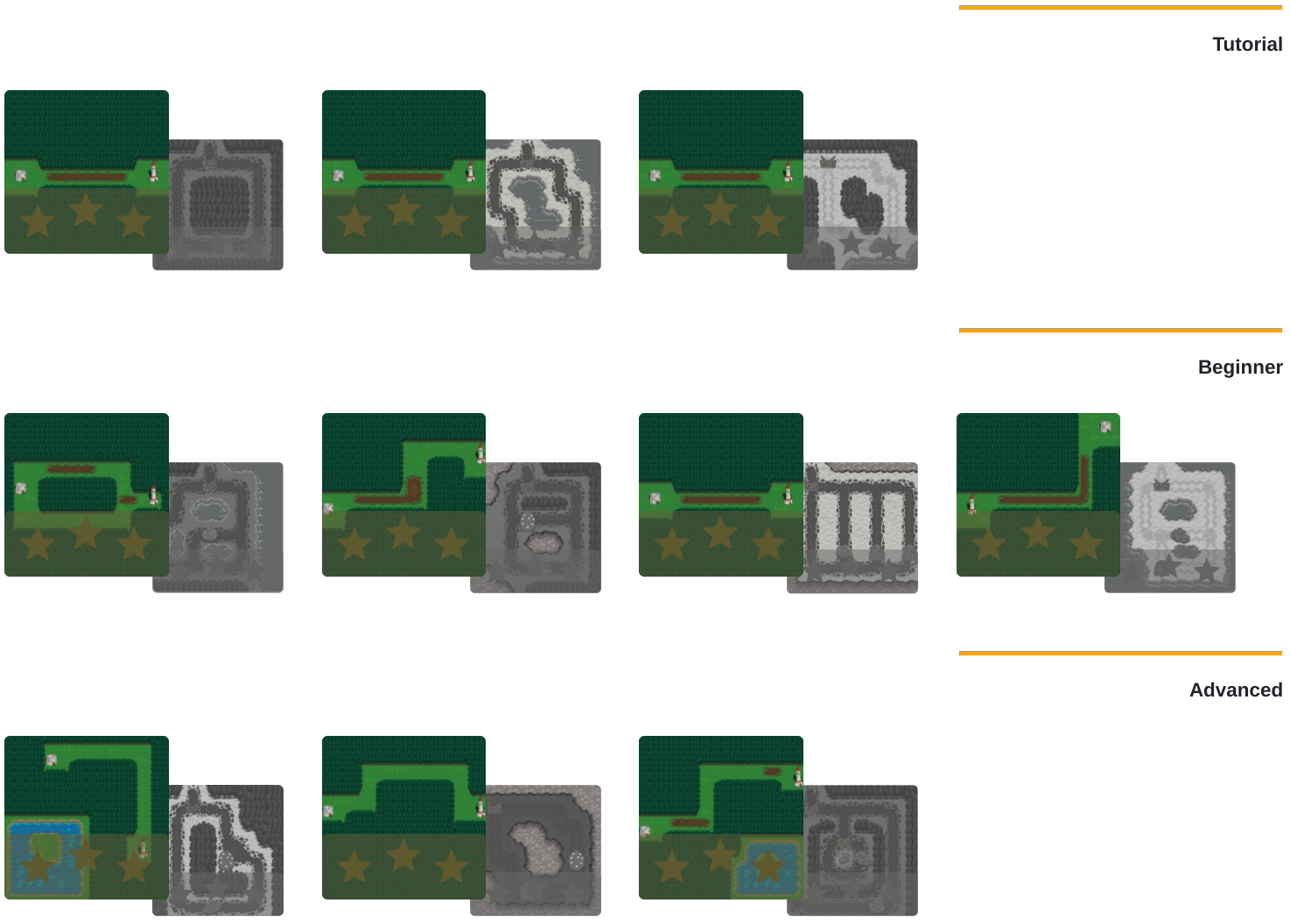}
	\caption{Level overview on the homepage with unlocked base and locked (grey) loop levels}
	\label{fig:allLevels}
\end{figure}

Together with the new loop levels, \toolname features a total of twenty levels: ten increasingly difficult base levels and ten corresponding second-stage levels. \Cref{fig:allLevels} displays the starting page of \toolname, organized into three difficulty categories, showcasing all twenty available levels.

\begin{figure*}
	\centering
	\includegraphics[width=0.8\linewidth]{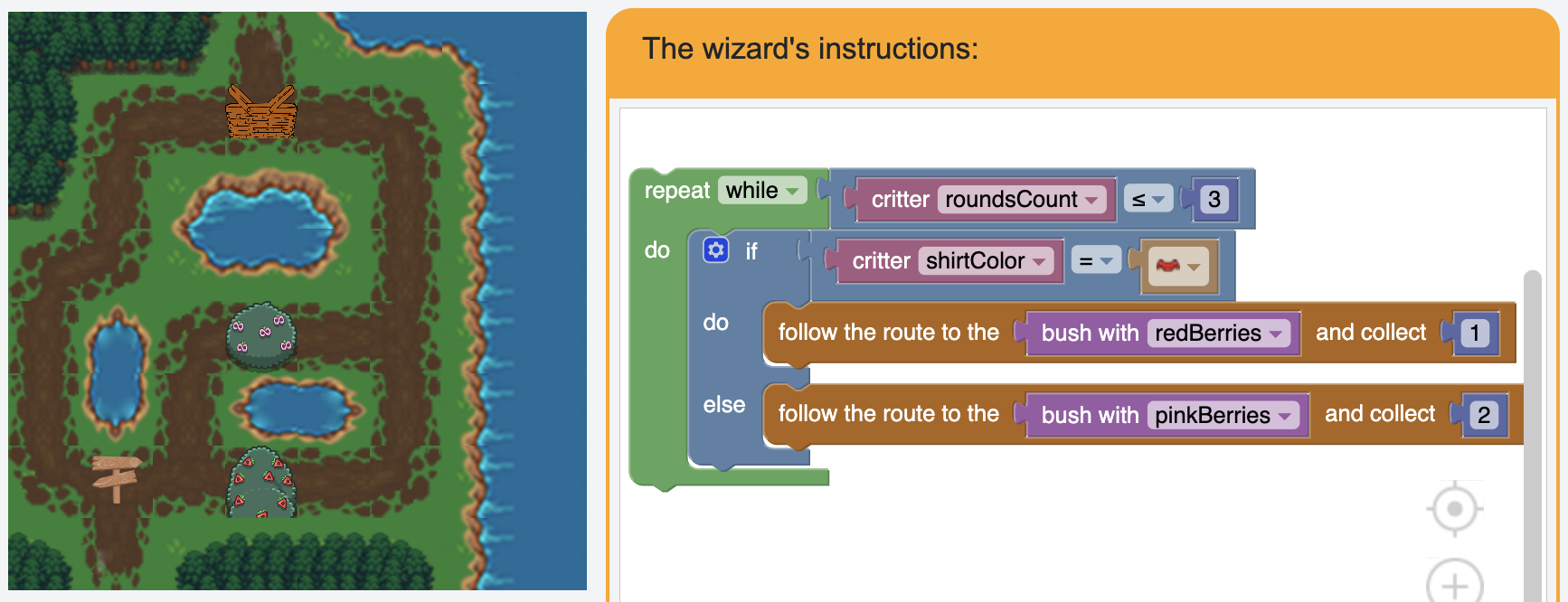}
	\caption{Gameboard and instructions for loop level 2}
	\label{fig:otherLoopLevel}
\end{figure*}

As players advance through the loop levels, both the Recipes Under Test and the game boards become more complex to increase the challenge. For example, \cref{fig:otherLoopLevel} illustrates a more advanced level that introduces a second type of berry and incorporates an if-condition related to the critters' shirt color into the recipe. The game board also includes an alternative route, visually representing this if-condition.

\begin{figure}
	\centering
	\includegraphics[width=\linewidth]{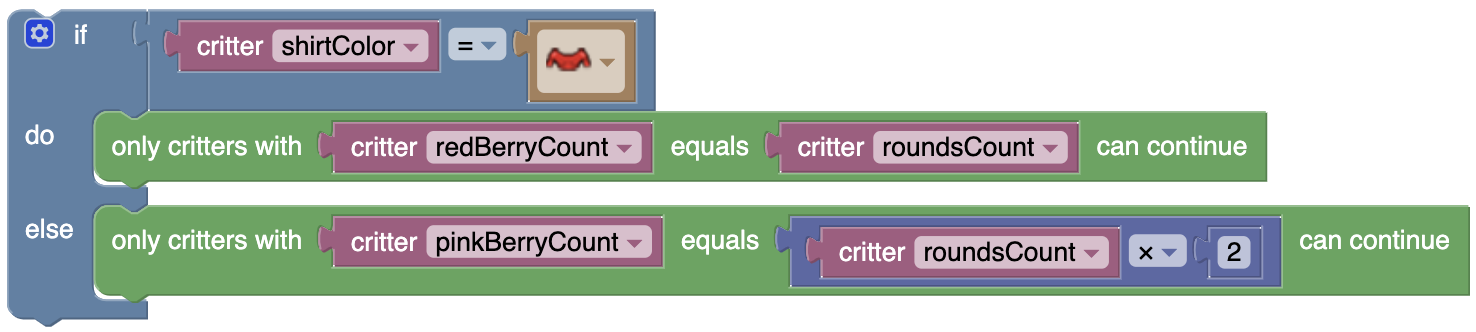}
	\caption{Correct test for loop level in \cref{fig:otherLoopLevel}}
	\label{fig:otherLoopLevelTest}
\end{figure}

\Cref{fig:otherLoopLevelTest} shows an example of a correct test, where the test ensures that the appropriate amount of either red or pink berries is collected, depending on the critter's shirt color.

As the game progresses, the complexity of the RUTs and game boards gradually increases, leading up to the final loop level. This gradual increase in difficulty helps to reinforce the fundamentals of loops and loop testing. The final loop level introduces the concept of nested loops, one of the most challenging loop structures to understand and test.


		\section{Evaluation}
		To assess the effectiveness of the new loop levels in \toolname, we conducted a controlled experiment focused on addressing the following research questions:

\begin{itemize}
	\item \textbf{RQ 1}: How do children play \toolname?
	\item \textbf{RQ 2}: How do children interact with the new loop-based levels in \toolname?
	\item \textbf{RQ 3}: Do children enjoy playing the loop levels of \toolname?
\end{itemize}

\subsection{Experiment Setup}

The controlled experiment was carried out at \school in two sessions in July 2024.

\subsubsection{Experiment Environment}

Through a collaboration with \school, a secondary school that prepares students for higher education, we conducted our experiment in their academic environment. We were given access to two sessions of their elective robotics course, each lasting 90 minutes and attended by around 15 students each. Since this course is elective, all students had shown interest in programming. Before the experiment, these students had ten months of experience with a block-based programming language developed by Lego,\footnote{\url{https://education.lego.com/en-gb/lessons/ev3-robot-trainer/}} which provided them with a solid understanding of block-based programming—the foundation of \toolname.

We had already used the original version of \toolname~\cite{DBLP:conf/icst/StraubingerBF24} during a prior session at the school, meaning that all the students were already familiar with the base levels. To avoid overwhelming the students in the limited time available, five base levels (1, 2, 6, 7, and 10) and five second-stage levels (1, 4, 6, 8, and 10) were selected. A total of 29 students participated in our experiment, conducted across two sessions. All participants were male, with no female students involved. The majority were in 5th to 7th grade (ages 11–13), with only three students from the 8th grade (ages 14–15). The experiment was organized in the school's computer lab, with each student working on their own computer.

\subsubsection{Experiment Procedure}

During the first ten minutes, we provided a recap of \toolname, covering its storyline, game mechanics, and the specifics of the new loop levels.
After the introduction, participants were tasked to play \toolname independently, without guidance on which level to choose, whether they should collect all points before moving on, or focusing on loop levels only. This gameplay session lasted for 60 minutes and ended with the conclusion of the experiment, providing enough time for participants to complete our exit survey. The survey starts with general questions about participants' gender and grade level, followed by seven questions on a five-point Likert scale that assess their enjoyment of various aspects of \toolname. It also includes open-ended prompts for participants to share additional thoughts or feedback.

\subsubsection{Experiment Analysis}

Our analysis centers on presenting results for all participants, highlighting the differences between the base and loop levels, as well as the challenges faced.

\subsubsection{RQ 1: How do children play \toolname?}

To answer this research question, we focus on understanding how children interact with the game. This analysis will help us identify the strategies and behaviors they use when playing the different levels, especially the newly introduced loop levels. We examine the data gathered from the experiment, covering both the base and loop levels, to gain these insights. First, we look at the average number of (1) completed and (2) attempted levels. Next, we examine the participants' activity throughout the experiment, focusing on any behavioral differences between base and loop levels.
This includes tracking which levels they played at various times and how many games they played during specific time intervals. We also perform a comparative analysis of three key metrics: the total number of (1) generated test cases (i.e., portals, signposts, and code blocks), (2) identified bugs (i.e., mutants and recipes), and (3) recognized correct code (i.e., healthy critters and collectors). This data is analyzed separately for the base and loop levels and is presented using box plots.

\begin{figure*}
	\vspace{-1.5em}
	\centering
	\begin{subfigure}[t]{0.325\textwidth}
		\centering
		\includegraphics[width=\textwidth]{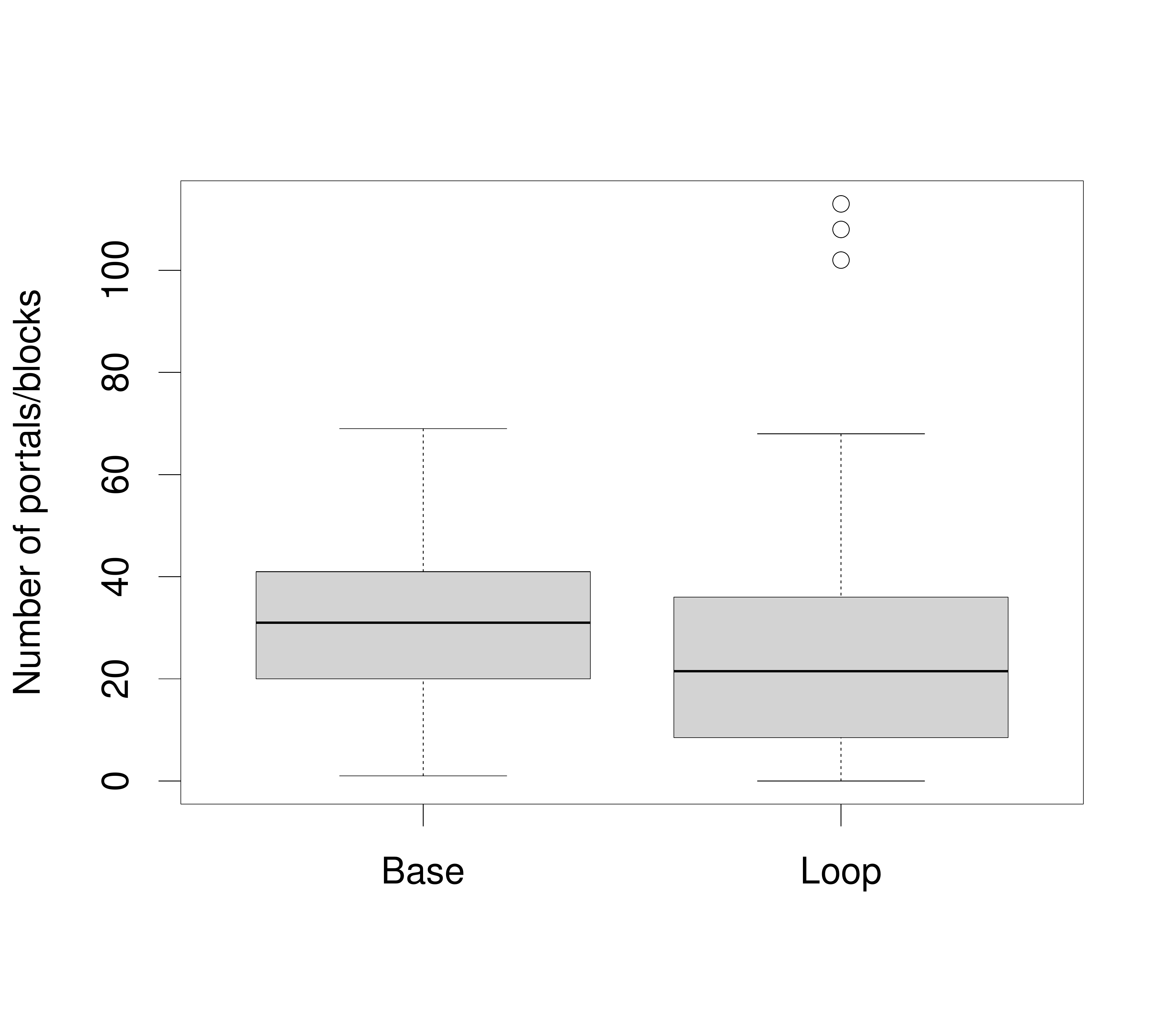}
		\vspace{-3.5em}
		\caption{Number of created portals/blocks}
		\label{fig:boxmines}
	\end{subfigure}
	\hfill
	\begin{subfigure}[t]{0.325\textwidth}
		\centering
		\includegraphics[width=\textwidth]{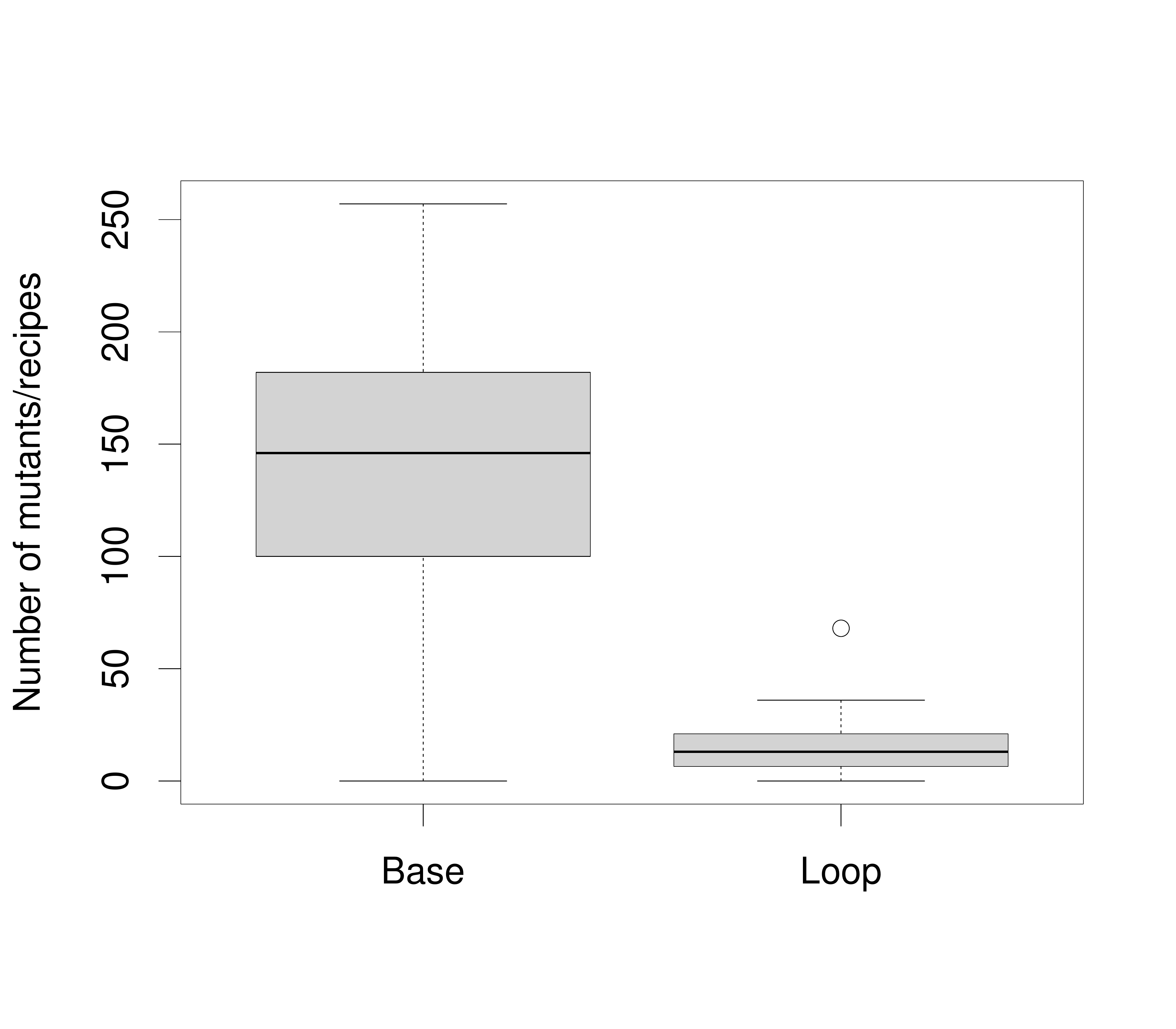}
		\vspace{-3.5em}
		\caption{Number of killed mutants/recipes}
		\label{fig:boxmutants}
	\end{subfigure}
	\hfill
	\begin{subfigure}[t]{0.325\textwidth}
		\centering
		\includegraphics[width=\textwidth]{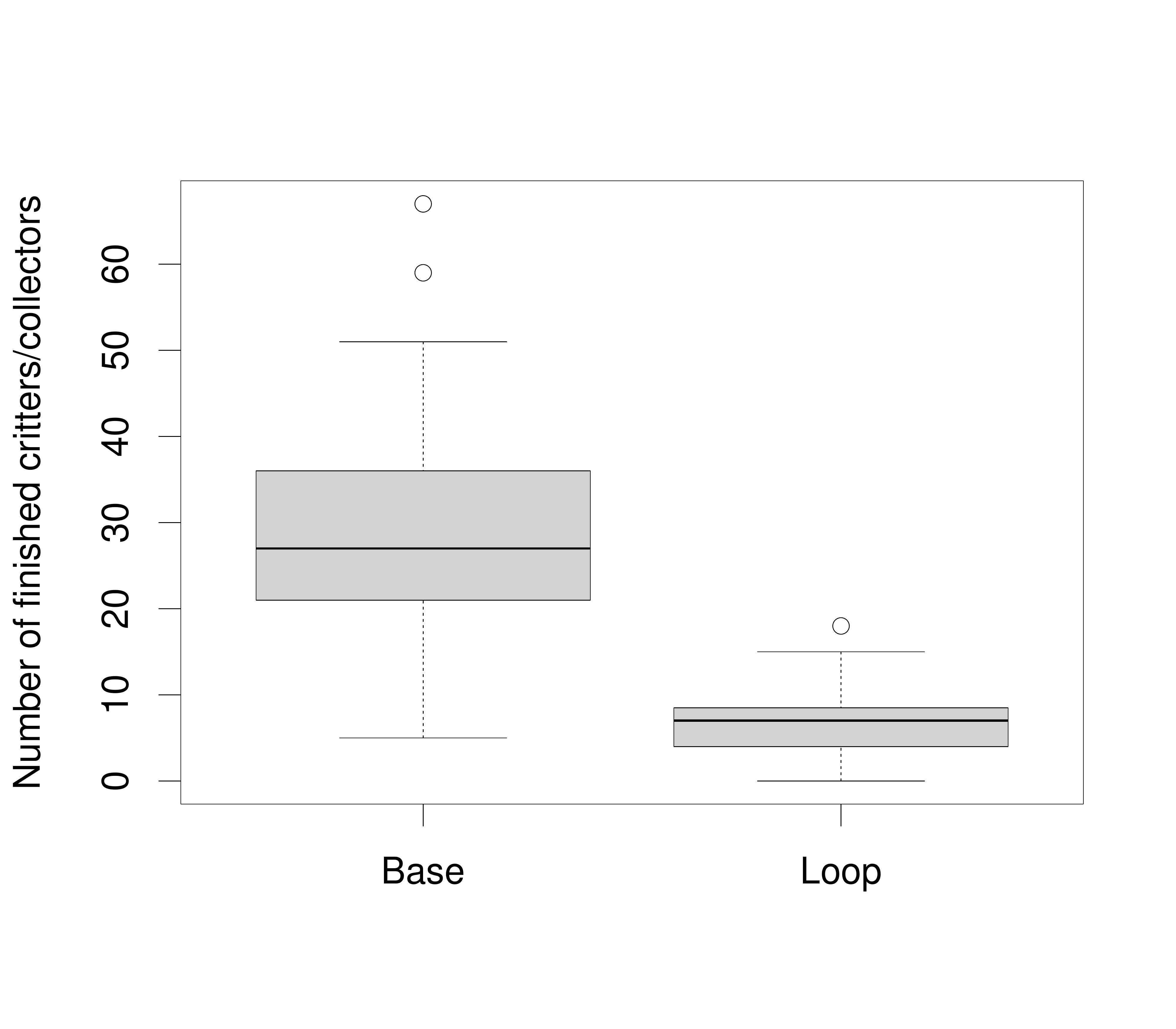}
		\vspace{-3.5em}
		\caption{Number of finished critters/collectors}
		\label{fig:boxhumans}
	\end{subfigure}
	
	\caption{Statistics on the use of \toolname divided into base and loop levels}
	\label{fig:testboxes}
\end{figure*}

\subsubsection{RQ 2: How do children interact with the new loop-based levels in \toolname?}

To answer this research question, we aim to assess how the children interact with the new loop levels of \toolname. Analyzing their behavior will help us understand their grasp of key concepts like loops and conditional logic while testing code. This insight will also highlight areas where they face challenges, guiding potential improvements in the game's design to enhance their learning experience. Specifically, we analyze three key ratios: (1) the number of utilized versus required blocks in signposts to assess their ability to express loop invariants, (2) the identified errors in recipes as a measure of their success, and (3) the correctly identified recipes. We examine these ratios across levels and over time to identify variations in difficulty and to track the skill development throughout the experiment.



\subsubsection{RQ 3: Do children enjoy playing the loop levels of \toolname?}

By examining their feedback and in-game behavior, we can determine which elements of the loop levels appeal to the children and which areas may need improvement to enhance the gaming experience. This insight will guide us in refining the game's design to make the learning process more engaging and effective. We analyze the responses from the exit survey, presenting the data through stacked bar charts that display the questions and their respective percentages.

\subsection{Threats to Validity} 

\paragraph{Threats to Internal Validity} Participants' previous experience with block-based programming and \toolname may influence the outcomes of the experiment. Introducing loop levels in \toolname without prior exposure to block-based programming could overwhelm some children. Additionally, participants might feel pressured to provide socially desirable responses in the exit survey, which could distort the data. To mitigate this risk, we encouraged them to answer honestly and without hesitation.

\paragraph{Threats to External Validity} The small sample size and lack of diversity in gender and grade levels restrict the ability to generalize our findings to a broader population of children. Since the participants were already enrolled in a robotics course and had an interest in programming, the results may not accurately reflect the experiences of children in more typical settings. Furthermore, the brief duration of the experiment may not capture long-term effects or usage patterns of \toolname, which could influence how interactions with the tool evolve over time.


		\section{Results}
		\subsection{RQ 1: How do children play \toolname?}

\begin{figure*}[t]
	\centering
	\begin{subfigure}[t]{0.45\textwidth}
		\centering
		\includegraphics[width=\textwidth]{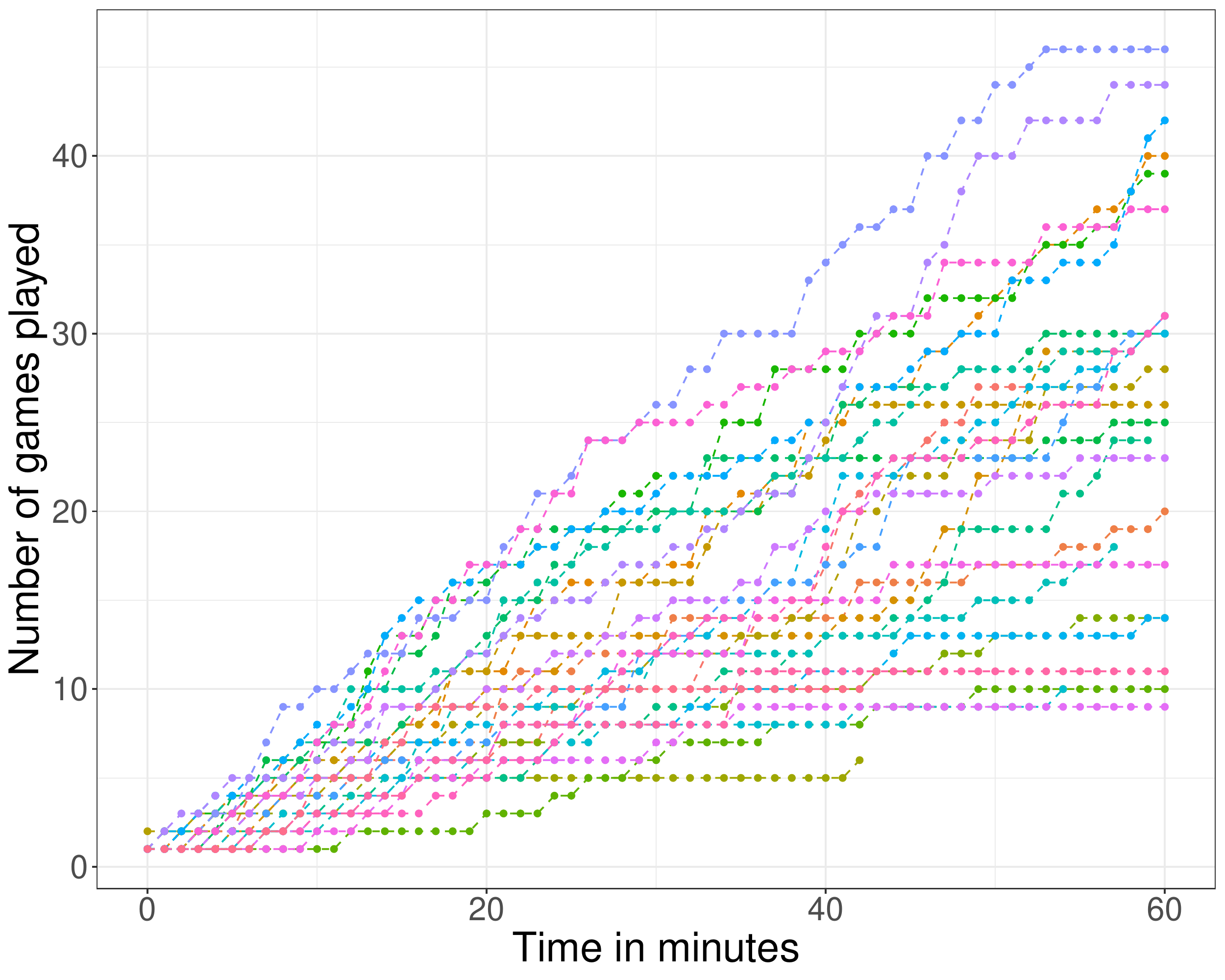}
		\caption{Number of games played over time}
		\label{fig:timegames}
	\end{subfigure}
	\hfill
	\begin{subfigure}[t]{0.45\textwidth}
		\centering
		\includegraphics[width=\textwidth]{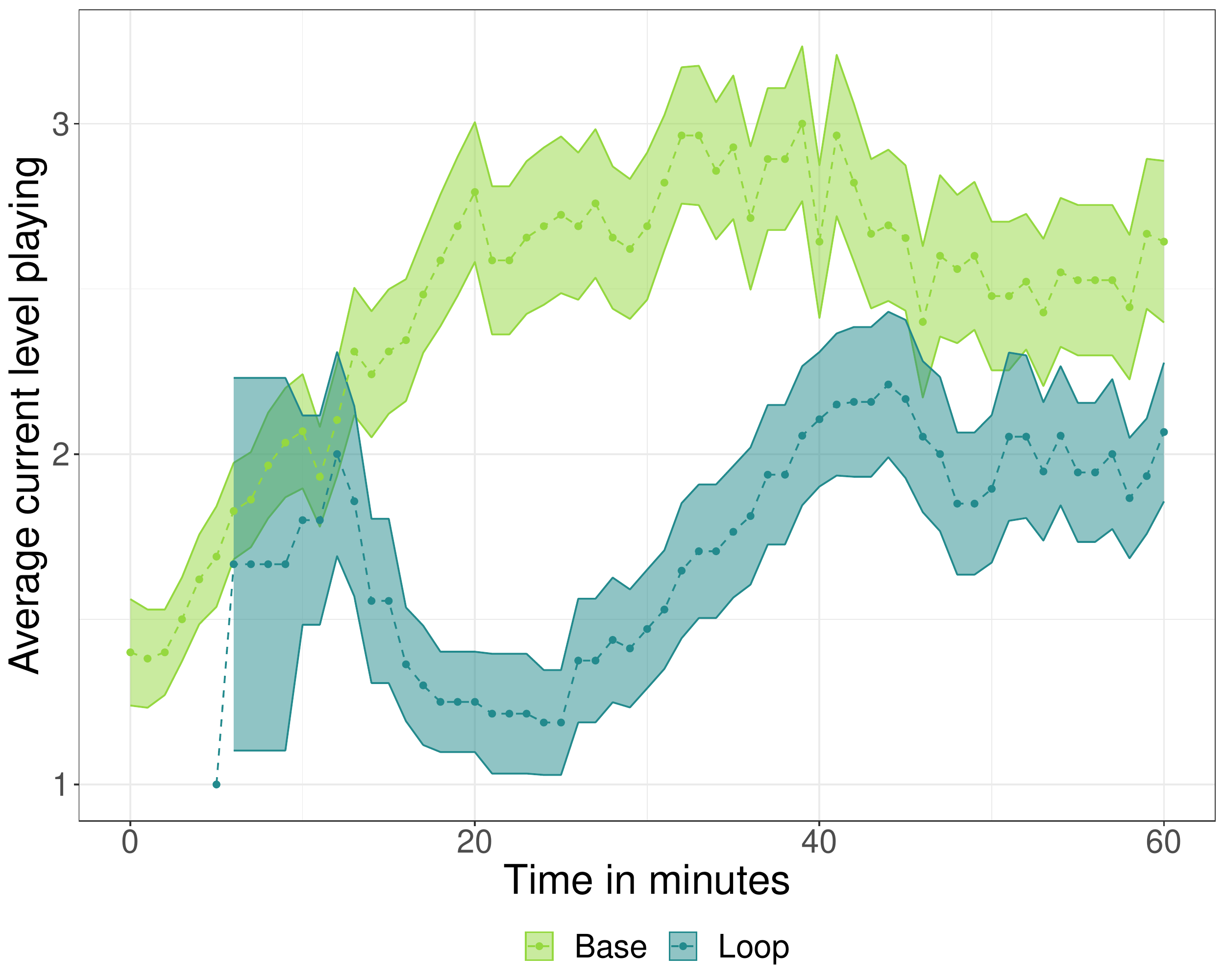}
		\caption{Current level played over time}
		\label{fig:timelevel}
	\end{subfigure}
	
	\caption{Differences between the players over time divided into base and loop levels}
	\label{fig:difftime}
\end{figure*}

During the experiment, the children played a total of 533 base-level games and 191 loop-level games, averaging 18.38 base-level games and 7.96 loop-level games per child. On average, they explored 4.03 base levels and 2.71 loop levels, successfully completing 1.72 base levels and 1.00 loop levels with a score of at least two out of three stars.
To unlock the corresponding loop level, they needed to achieve at least two stars in the base level. On average, the children needed about three attempts to unlock a loop level, resulting in them playing the base levels three times each. Once the loop level was unlocked, players often attempted it immediately but would return to the base level to maximize their score if they could not make quick progress. Therefore, the children performed better on base levels due to familiarity, while loop levels posed greater challenges, prompting repeated attempts.

During gameplay, the children placed an average of 31.45 portals—most of the time consisting of three blocks each—in the base levels and utilized 32.15 blocks in the loop level signposts (\cref{fig:boxmines}).
They also detected an average of 137.31 mutants in the base levels and identified 16.71 faulty recipes in the loop levels (\cref{fig:boxmutants}). Additionally, the players allowed 29.24 healthy critters to reach the tower in the base levels, and on average, 6.58 collectors successfully gathered the correct amount of berries in the loop levels (\cref{fig:boxhumans}). These numbers vary significantly because (1) the children played more base than loop levels, and (2) the base levels feature more critters on the gameboard compared to the loop levels. Additionally, in the base levels, players could place and remove portals freely along the path, allowing for more flexible experimentation. In contrast, blocks in the loop levels could only be set at specific signposts, which required more precise decision-making and offered a wider variety of blocks for testing. This makes it challenging, if not impossible, to directly compare the base levels with the loop levels, which is why we refrain from doing so and present the data as it is.

When examining the number of games played over time (see \cref{fig:timegames}), it is clear that there are significant differences among the children. Some played as few as seven games in total, while others played up to 46 games during the 60-minute experiment. This wide range suggests that the children either have varying levels of programming skills or that some were not fully engaged in the activity. Observations during the experiment indicated the latter in some cases, as a few children attempted to cheat by modifying blocks and scores using the developer consoles in their browsers. This occurred only a few times, and altering the data in the browser does not affect any information in the database.
Nevertheless, they interacted with \toolname in one way or another. The teacher also noted that the children seemed more engaged with \toolname than they usually are with their typical tasks, where they tend to get distracted more often.

\cref{fig:timelevel} illustrates the progression of levels played, showing the current level (base or loop) that each player opened at any given minute during the experiment on average for all players. Initially, only the base levels were played since the loop levels were locked. After a few minutes, when the first loop levels were unlocked, many children attempted to play either loop level one or two. However, many quickly returned to loop level one to first understand how these loop levels worked. The progression through base levels increased steadily during the first third of the experiment, then remained mostly stable for the rest of the session, with a slight decline toward the end. Meanwhile, the number of loop levels played increased significantly in the second third of the experiment as more children managed to unlock them, reaching a peak around the 40-minute mark and stabilizing afterward, with most children playing loop levels one through three.

\summary{RQ 1}{Our study reveals wide variations in gameplay patterns, level progression, and involvement, with some even attempting to manipulate the game. Despite these differences, overall engagement was higher than in their usual activities, with most children gradually advancing through the levels after an initial learning phase.}

\subsection{RQ 2: How do children interact with the new loop-based levels in \toolname?}

\begin{figure*}
	\centering
	\begin{subfigure}[t]{0.45\textwidth}
		\centering
		\includegraphics[width=\textwidth]{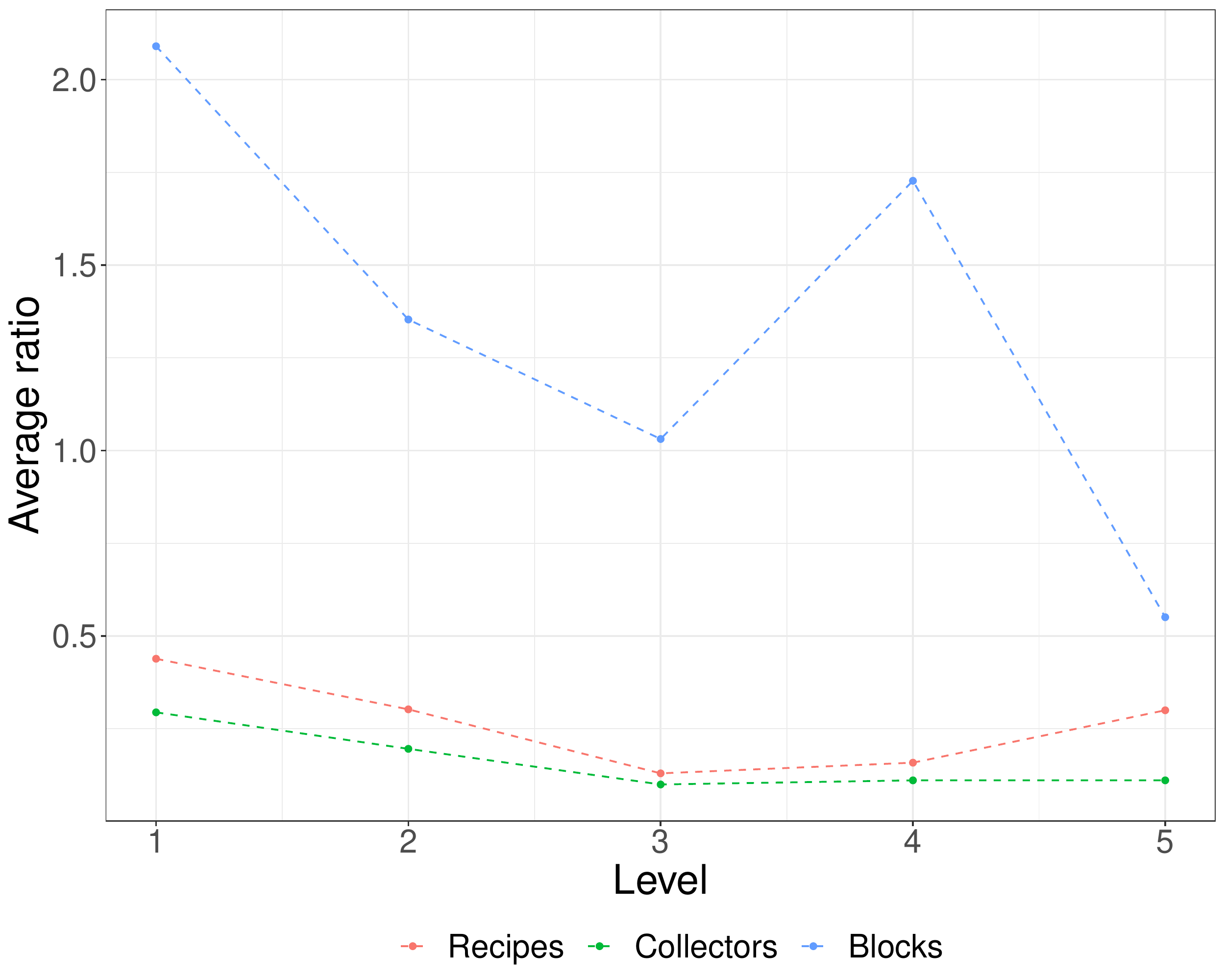}
		\caption{Correctness per loop level}
		\label{fig:correctnesslevel}
	\end{subfigure}
	\hfill
	\begin{subfigure}[t]{0.45\textwidth}
		\centering
		\includegraphics[width=\textwidth]{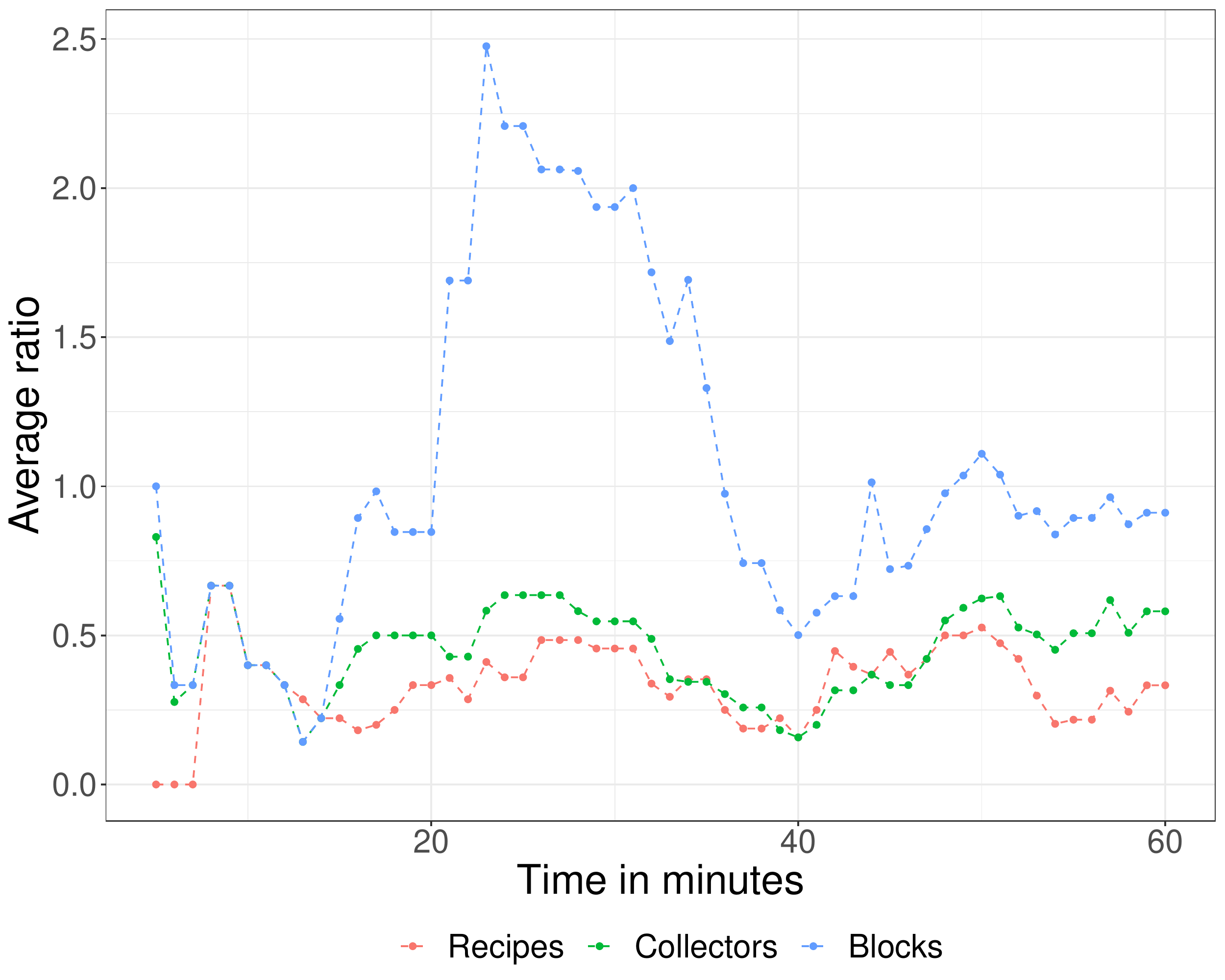}
		\caption{Correctness over time for loop levels}
		\label{fig:correctnessratio}
	\end{subfigure}
	
	\caption{Differences in correctness on level and time basis}
	\label{fig:diffcorrecttime}
	\vspace{-1em}
\end{figure*}

\begin{figure}
	\centering
	\includegraphics[width=\linewidth]{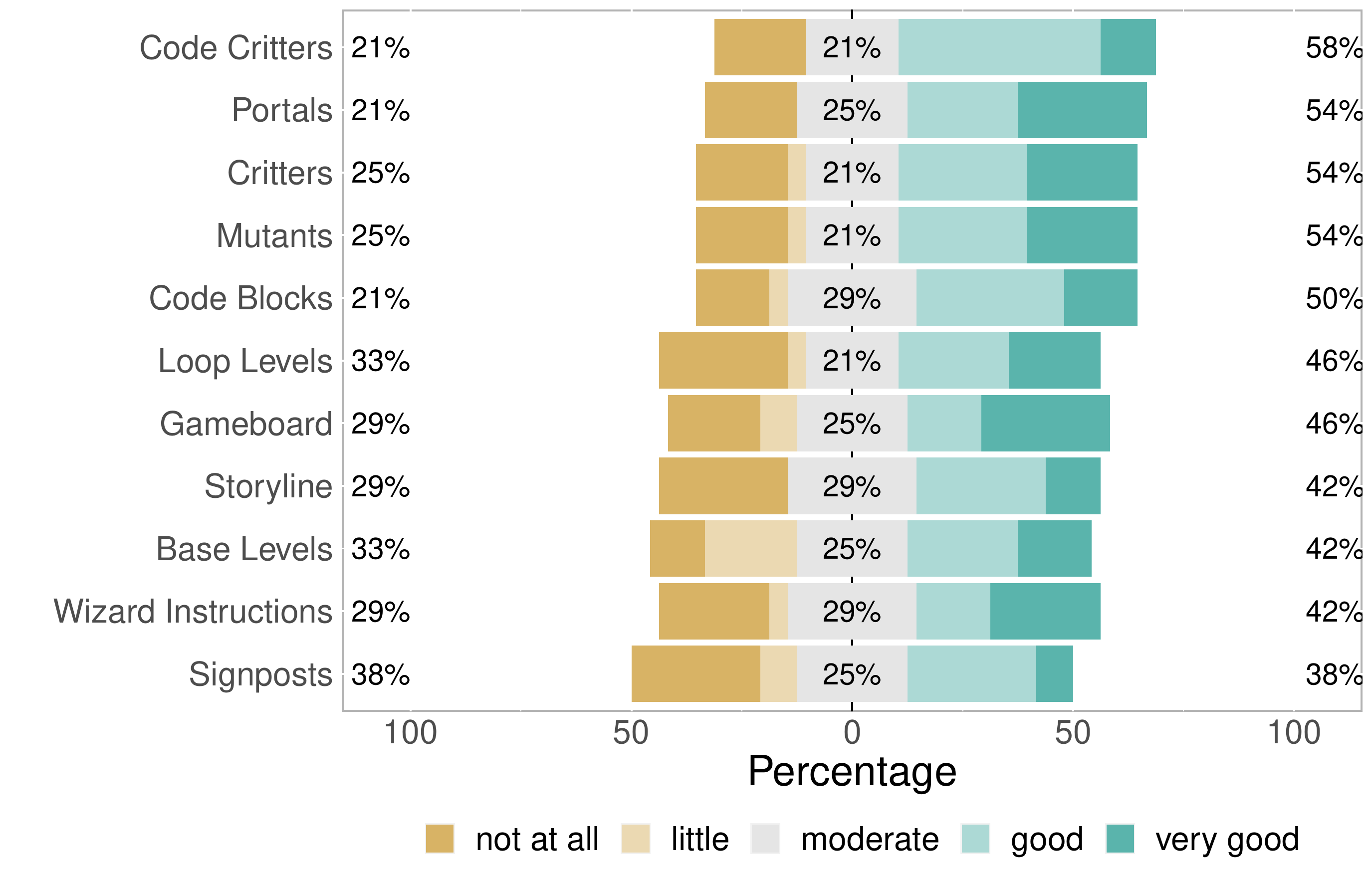}
	\caption{Survey responses to ``How much did you enjoy\ldots''}
	\label{fig:survey}
\end{figure}

\Cref{fig:timelevel} shows that most players concentrated on loop levels 1 and 2, with only a few attempting levels 3 and 4 before returning to the earlier levels, which accounts for the significantly lower ratios in levels 3 and 4. Even fewer students tackled level 5, but those who did were more successful at identifying incorrect recipes than in levels 3 and 4 (\cref{fig:correctnesslevel}), despite level 5 being the most challenging due to its introduction of nested loops. This suggests that only the strongest players attempted level 5, leading to higher average success rates. \cref{fig:correctnesslevel} shows that the ratios of incorrect recipes and completed collectors remain fairly consistent in their differences across each level. Across all levels, the ratio of detected incorrect recipes was consistently higher than that of completed collectors, indicating a focus on identifying code errors over recognizing correct code. Although both are connected since they travel the same path, many children wrote tests that correctly identified all the incorrect recipes but also mistakenly caught the collectors due to flawed testing. During this testing, the number of blocks used in level 1 was more than twice the optimal amount, resulting in a ratio of more than one, but this number decreased in levels 2 and 3, likely reflecting the players' learning curve. The increase in block usage for level 4 in \cref{fig:correctnesslevel} is due to multiple if-conditions within the loop, requiring more blocks to handle its complexity.

Looking at the average ratios over time in \cref{fig:correctnessratio}, a different pattern emerges. Initially, the ratio of correctly identified recipes (collectors in \cref{fig:correctnessratio}) is near its maximum, while the ratio of detected incorrect recipes (recipes in \cref{fig:correctnessratio}) starts at zero. This suggests that players initially allowed the levels to run without adding tests to familiarize themselves with the loop levels. As the experiment progressed, the ratios for both recipes and collectors fluctuated similarly between 0.3 and 0.6, with a noticeable drop around two-thirds of the way through the experiment. This decline might indicate that players transitioned from the first to the second level, requiring time to adjust to the increased difficulty. After this dip, both ratios improved, showing that players adapted and learned how to handle the new level. Towards the end, the ratio of detected incorrect recipes decreased again, possibly indicating that some children attempted new levels but could not make further progress as time ran out. The peak in the ratio of blocks added during the second third of the experiment indicates that many players initially attempted to use as many blocks as possible to cover all scenarios, often exceeding a ratio of one, which would represent the optimal number of blocks. This approach was gradually abandoned, likely because they did not see a corresponding increase in accuracy, as reflected in the steady ratios for both recipes and collectors. This indicates that the students learned that including too many assertions in a single test can create confusion rather than improve results, making it difficult to identify what went wrong.

\summary{RQ 2}{Children playing the new loop-based levels were better at identifying errors in code than recognizing correct code, with most focusing on the simpler levels and adapting their strategies as difficulty increased.}

\subsection{RQ 3: Do children enjoy playing the loop levels of \toolname?}

\Cref{fig:survey} summarizes the results of the exit survey, showing that 58\% of the children enjoyed playing \toolname. While they liked the elements of the base levels—such as portals, critters, and mutants—they were more uncertain about the loop levels, particularly the wizards' instructions and the signposts. The signposts of the loop levels were less liked by the players than other features, with 38\% of players expressing dissatisfaction, a sentiment frequently noted in the free-text responses.  It was not that they disliked the loop levels more than the base levels; instead, they wished the portals they enjoyed in the base levels could also be used in the loop levels, as they were already familiar with them. Post experiment, we addressed this feedback by replacing the signposts with movable portals like those in the base levels in a newer version of \toolname.  Conversely, many children expressed their enjoyment of playing \toolname in the free-text comments, stating that they liked it and encouraged further development of the game with additional levels and stories. Notably, two participants even logged into \toolname from home. 

\summary{RQ 3}{The children enjoyed playing \toolname, favoring the base levels over the loop levels due to their familiarity with the portals.}

		\section{Discussion}
		The experiment revealed several challenges that children faced while interacting with \toolname, which varied by game level and were especially noticeable with advanced programming concepts like loops and conditionals.

A major hurdle was understanding loops and how they were represented in the game. Many participants struggled to connect loop iterations (rounds) with the expected outcomes (like berry collection per round). In level 1, most players initially misunderstood how loops worked, leading to unsuccessful attempts at designing tests to distinguish mutants from valid collectors. Though they often figured it out after two or three tries, the process highlighted a significant learning curve. Despite nearly a year of programming lessons, many students lacked a clear grasp of fundamental loop concepts, indicating that their learning focused more on using loops in a limited setup rather than understanding the underlying principles.

\begin{figure}
	\centering
	\includegraphics[width=\linewidth]{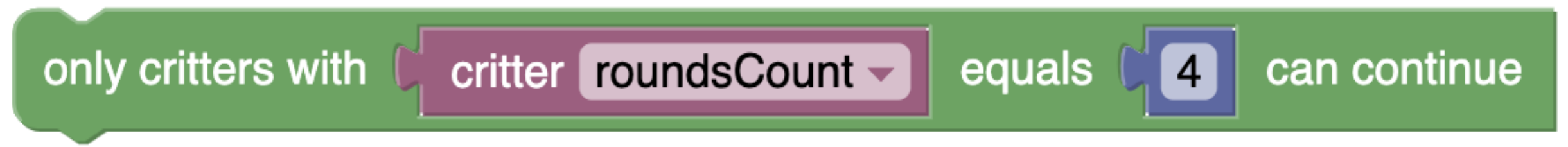}
	\caption{Incorrect test for loop level 1}
	\label{fig:invalidTest1}
\end{figure}

\begin{figure}
	\centering
	\includegraphics[width=\linewidth]{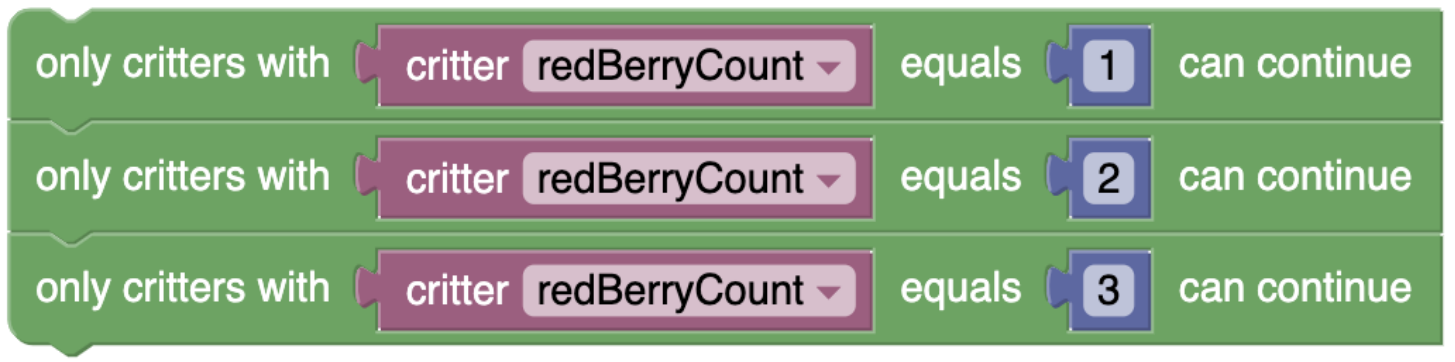}
	\caption{Incorrect test for loop level 1}
	\label{fig:invalidTest2}
\end{figure}

One example of an incorrect test for level 1 is shown in \cref{fig:invalidTest1}. Since the wizard's instructions specify that the critters should walk three rounds, the player was likely trying to prevent them from walking a fourth round. However, the way the test was set up resulted in the opposite effect: all critters were sent back in the first round because their \texttt{roundsCount} was not equal to four. Another example is the test shown in \cref{fig:invalidTest2}. This test contains three assertions that compare the current number of collected red berries to one, two, and three. Although it seems that the player grasped the principles of loops, they misunderstood that all assertions are evaluated during each pass of a critter. As a result, this leads to all recipes being marked as incorrect, since the number of berries cannot simultaneously be one, two, and three.

Level 2 introduced if-conditions within loops, which significantly increases the complexity. Most students found it difficult to apply conditional logic to differentiate between critters based on their attributes. This difficulty was amplified by the need to evaluate these conditions within the loop repeatedly, resulting in frequent mistakes. Only a small fraction of the participants (four out of 20) completed level 2 successfully, underscoring the lack of foundational programming knowledge—a gap that \toolname was not originally designed to address.
The introduction of nested loops in later levels added another layer of complexity, requiring participants to manage multiple variables over several iterations. This abstract reasoning was a steep challenge, and no participants fully mastered these levels.
While most students engaged well with the base levels, their interest dropped with the advanced loop levels. Some even attempted to alter the game's code using developer tools, possibly out of frustration with the game mechanics or their lack of progress. This behavior indicates a need to better support motivation and understanding, especially when the difficulty increases.

To help with these challenges, future versions of \toolname could introduce loops and conditionals more gradually, emphasizing the difference between individual iterations and overall loop behavior. Breaking down complex concepts and incorporating tutorials could strengthen foundational skills before progressing to harder levels. Adding an in-game hint system could also provide real-time support, guiding players through loops and conditional logic. Hints suggesting ways to structure tests or manage variables could help refine their approach, reducing frustration and maintaining engagement.

		\section{Conclusions}
		
The \toolname game demonstrated~\cite{DBLP:conf/icst/StraubingerBF24} that even young learners can engage
with testing concepts in an engaging and fun way, but the game design
precludes the important programming concept of loops. To address this
problem, we introduced a new game concept with loop-based recipes to
teach children about loops. The narrative of our game extends the one
introduced by \toolname and provides a satisfying story of healing
the infected critters collected as part of the original gameplay. The
gameplay uses the established approach of building on mutation testing
to introduce a testing challenge. Importantly, though, the concept of
recipes allows loops to become an integral aspect of the game. Our
study with 29 secondary school students confirms that the
loop-integrated levels promote active engagement, although they also
present challenges in understanding loops and conditionals.

To address these challenges, there are many possible avenues for
future exploration. The game could more gradually introduce different concepts,
making a clear distinction between single iterations and full loop
behaviors. This step-by-step approach, coupled with tutorials and
guided assistance, would help children develop a stronger
understanding before moving on to more complex levels. Adding an
in-game hint system could provide real-time guidance, especially for
loops and conditionals, offering helpful tips on structuring tests and
handling variables to keep players engaged and reduce frustration. By
incorporating adjustable difficulty levels, particularly in the loop
stages, the game could better accommodate for children with different
skill levels, ensuring a smoother progression through increasingly
challenging concepts. Visual aids that clearly illustrate how abstract
variables interact, such as rounds in a loop, would help connect the
game's mechanics to programming concepts, making it easier for
children to grasp how loops work and their effects on gameplay. These
enhancements would make \toolname more intuitive and engaging,
ultimately supporting young learners as they master testing
fundamental programming concepts.
\vspace{1em}


		\noindent The source code of \toolname is available at:
		
		\begin{center}
			\href{https://github.com/se2p/code-critters}{https://github.com/se2p/code-critters}
		\end{center}
		
		\noindent You can give \toolname a try online at:
		
		\begin{center}
			\href{https://code-critters.org}{https://code-critters.org}
		\end{center}
	
		To support replications, all source code and experiment materials used in our study are available at:
		\begin{center}
			\url{https://doi.org/10.6084/m9.figshare.28343282}
		\end{center}
		
		\balance
		\bibliographystyle{IEEEtran}
		\bibliography{bib}
		
	\end{document}